\documentclass[showpacs,aps,graphicx,]{revtex4}
\usepackage{amsmath}
\usepackage{mathrsfs}
\usepackage{amsfonts}
\usepackage{amssymb}
\usepackage{graphicx}
\usepackage{caption}
\usepackage{eufrak}
\usepackage{multirow}
\usepackage{float}
\usepackage[colorlinks,linkcolor=blue,anchorcolor=blue,citecolor=blue]{hyperref}
\usepackage{soul,color,xcolor}
\soulregister\upcite7
\soulregister\citep7
\soulregister\citet7
\soulregister\ref7
\soulregister\pageref7
\soulregister\and7
\soulregister\Name7

\begin{document}

\title{Improved entanglement-based high-dimensional optical quantum computation with linear optics}

\author{Huan-Chao Gao\textsuperscript{1}, Guo-Zhu Song\textsuperscript{2}, and  Hai-Rui Wei\textsuperscript{1}\footnote{Corresponding author: hrwei@ustb.edu.cn} }

\address{1 School of Mathematics and Physics, University of Science and Technology Beijing, Beijing 100083, China \\
%
%
2 College of Physics and Materials Science, Tianjin Normal University, Tianjin 300387, China}

\date{\today }

\begin{abstract}

Quantum gates are the essential block for quantum computer. High-dimensional quantum gates exhibit remarkable advantages over their two-dimensional counterparts  for some quantum information processing tasks. Here we present a family of entanglement-based optical controlled-SWAP gates on $\mathbb{C}^{2}\otimes \mathbb{C}^{d}\otimes \mathbb{C}^{d}$. With the hybrid encoding, we encode the control qubits and target qudits in photonic polarization and spatial degrees of freedom, respectively.
The circuit is constructed using only $(2+3d)$ ($d\geq 2$) linear optics, beating an earlier result of 14 linear optics with $d=2$. The circuit depth 5 is much lower than an earlier result of 11 with $d=2$. Besides, the fidelity of the presented circuit can reach 99.4\%, and it is higher than the previous counterpart with $d=2$. Our scheme are constructed in a deterministic way without any borrowed ancillary photons or measurement-induced nonlinearities. Moreover, our approach allows $d>2$.


\end{abstract}


\maketitle

\newcommand{\upcite}[1]{\textsuperscript{\textsuperscript{\cite{#1}}}}

\section{Introduction}\label{sec1}

Quantum computers has the potential ability to solve certain intractable computational problems, such as searching unsorted databases,\upcite{Grover1,Grover2} efficient quantum simulation\upcite{quantum simulation} and factoring large integers,\upcite{Shor} much more efficiently than their existing classical counterparts.\upcite{book}
Universal gate sets are the core building block for quantum computer, and they also play a central role in quantum communication like entanglement purification and concentration,\upcite{gate-communication} entanglement generation.\upcite{Li-Tao1,Li-Tao2} Until now, quantum gates have been demonstrated in various types of two-level (qubit) physical platforms ranging from photons,\upcite{gate-photon1,gate-photon2} neutral atoms,\upcite{gate-atom1,gate-atom2} trapped ions,\textsuperscript{\upcite{gate-ion1,gate-ion2}} nuclear magnetic resonance (NMR) to artificial atoms.\upcite{NMR,gate-artifical-atom}  
Controlled-NOT (CNOT) gates are the primitive and most popular entangling universal gate in two dimension system. CNOT gates assisted with general single-qubit rottions are sufficient to implement any multiqubit quantum computation.\upcite{Li-Tao2,universal} 
Three-qubit Toffoli and Fredkin (controlled-SWAP) gates have broad applications in reversible computing,\upcite{reversible} quantum error correction,\upcite{correction1,correction2} fault tolerance,\upcite{tolerance} quantum arithmetic circuits,\upcite{arithmetic} and quantum algorithm.\upcite{algorithm} Moreover, the Toffoli or Fredkin gate combination with Hadamard gate forms a universal gate.\upcite{Toffoli-universal1,Toffoli-universal2,Toffoli-universal3} 
Successful experimental demonstrations of the Toffoli and Fredkin gate have been reported in various underlying physical systems.\upcite{realize-Toffoli,realize-Fredkin,auxiliary entanglement}

Compared to two-dimensional qubit, high-dimensional qudit provides a larger state space,\upcite{large-capacity} higher noise-resilience,\upcite{noise-resilience} richer fault tolerance,\upcite{qudit-fault-tolerance} and stronger violation of Bell's non-locality\upcite{Bell-locality} in quantum information processing. Thus, qudit system has strong potential for reducing quantum complexity,\upcite{Toffoli-universal2,Toffoli-universal3,qudit-complexity1,qudit-complexity2} simplifying experimental setup,\upcite{qudit-simplify-setup} and enhancing the  efficiency and accuracy of quantum computation and algorithm.\upcite{algorithm,qudit-efficiency1,qudit-efficiency2} Nowadays, qudit-based technologies and devices have been proposed in photons,\upcite{qudit-photon1,qudit-photon2,qudit-photon3,qudit-photon4} trapped ions,\upcite{qudit-ion1,qudit-ion2} superconductors,\upcite{qudit-superconductor1,qudit-superconductor2,qudit-superconductor3,qudit-superconductor4,qudit-superconductor5} and solid-state systems.\upcite{qudit-solid1,qudit-solid2} Among all these candidates, photons are the ideal natural carrier of high-dimensional quantum information due to their variety of qudit-like degrees of freedom (DOFs), including spatial modes,\upcite{spatial1,spatial2,spatial3,spatial4,spatial5} orbital angular-momentum (OAM),\upcite{orbital1,orbital2,orbital3} optical frequencies,\upcite{frequency1,frequency2} and time of arrival.\upcite{noise-resilience,time-bin1,time-bin2}
In the meantime, high-dimensional quantum information can be encoded in multiple degree of freedom, such as hyperentanglement encoded in three DOFs system,\upcite{three DOFs system} polarization-spatial-mode hyperentanglement.\upcite{polarization-spatial1,polarization-spatial2}  Among all these DOFs, polarization and spatial DOFs are the most popular, because they can be extremely fast and accurate manipulated only with linear optics.

Despite significant progress has been made in qudit-based quantum information processing, there is much less research on qudit-based quantum gate. Some library of fundamental single-qudit gate, such as Clifford group, has been proposed. Spatial-based and OAM-based single-qudit gates were reported in photonic systems.\upcite{qudit-superconductor2,qudit-single1} Basic and primitive two-qudit gates are rarely investigated both theoretically and experimentally. Up to now, only controlled-SUM gate,\upcite{sum-wang,qudit-superconductor3,qudit-superconductor4,sum-Su} controlled extensions of rotation gates $\text{C}_m[R_d]$,\upcite{Controlled-rotation-Luo} generalized XOR (GXOR) gate,\upcite{qudit-ion1} controlled increment (CINC) gate,\upcite{qudit-ion2} generalized controlled X (GCX) gate\upcite{GCX-Di1,GCX-Di2} were proposed and extensively used. The research on those universal gates mainly focused on entangling power. Note that the synthesis of qudit-based quantum computing in terms of CINC and GCX was constricted recently.\upcite{qudit-ion2,GCX-Di1,GCX-Di2} Controlled-SUM gate was experimentally realized  in superconducting\upcite{qudit-superconductor4,sum-Su} and photonic architectures. In case of multi-qudit gate, the controlled-SWAP gate was experimental demonstration in $\mathbb{C}^{2}\otimes \mathbb{C}^{4}$ with polarization and OAM degrees of freedom of a single photon  in 2021.\upcite{CSWAP-Wang}
Recently, Liu et al.\upcite{high-dimensional quantum gates} demonstrated  high-dimensional quantum gates with ultrahigh fidelity and efficiency. It is a major advance in the development of high-dimensional quantum gates.

In this paper, we present a general scheme for implementing deterministic entanglement-based controlled-controlled SWAP (Fredkin) gate in $\mathbb{C}^{2}\otimes \mathbb{C}^{d}\otimes \mathbb{C}^{d}$ by using linear optics, that are not based on sequences of CNOT gates. Our work is inspired by recent work Ref. \cite{Fredkin-Meng}, which shows that three-qubit controlled-SWAP gate acting on polarization and spatial DOFs of the entangled photon pairs. By slightly modifying the spatial-encoding,
we first slit the depth of the quantum circuit for three-qubit controlled-SWAP gate from 11 to 5, and reduce the number of linear optics contained from 14 to 8. And then we extend such an approach to implement an controlled-SWAP gate in $\mathbb{C}^{2}\otimes \mathbb{C}^{d} \otimes \mathbb{C}^{d}$ ($d\geq3$) using only $2+3d$ linear optics, and the $d$-independent the circuit shadow is 5.
Moreover, the fidelity of our scheme is over the counterpart shown in Ref. \cite{Fredkin-Meng}.


\section{Implementation of a controlled-SWAP gate $U_{\text{CSWAP}}^{2\otimes2\otimes2}$} \label{sec2}

\begin{figure}[htbp]
\centering
\includegraphics[width=12 cm]{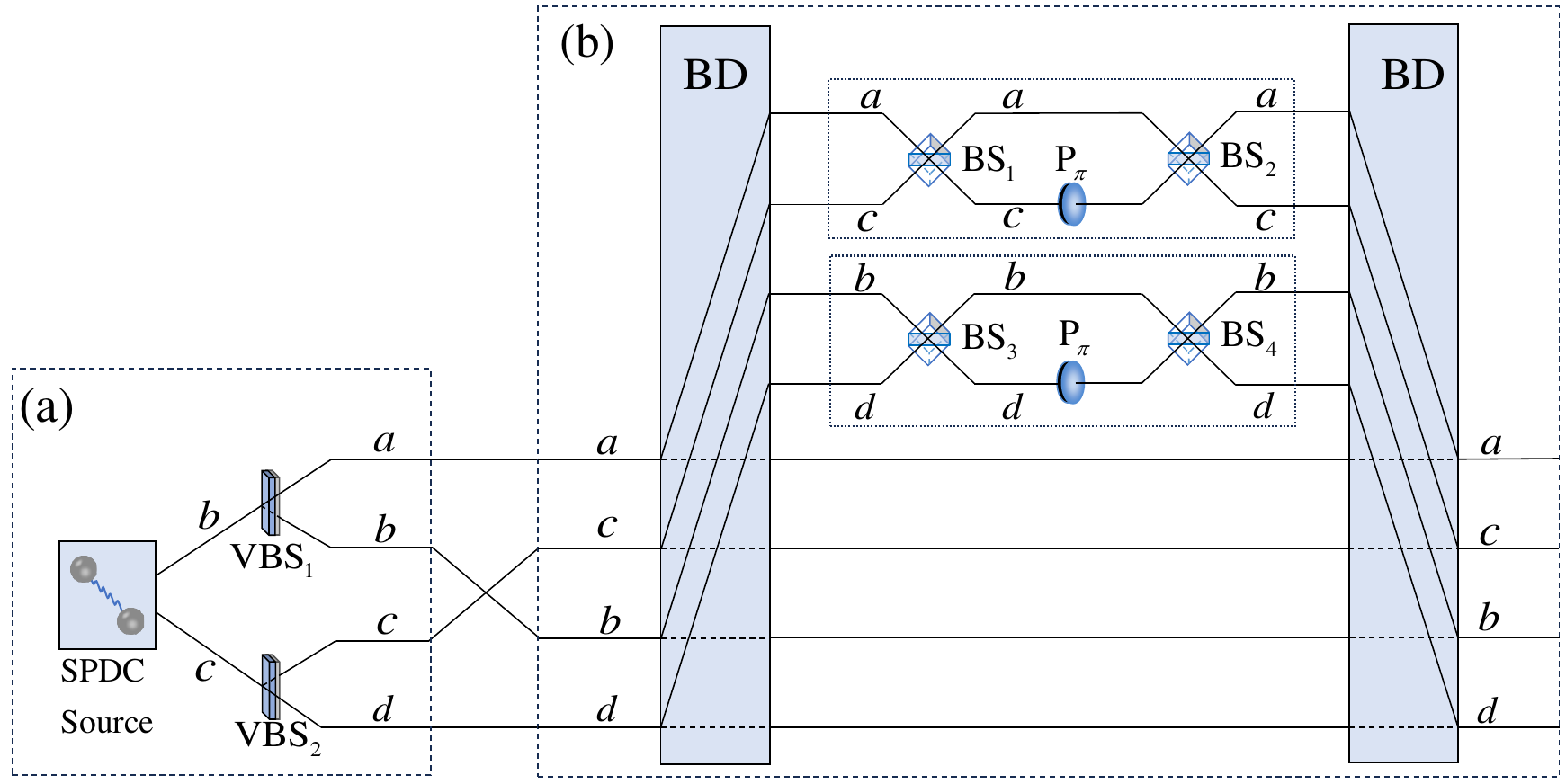}
\caption{a)
Schematic diagram of the device for state-preparation of controlled-SWAP gate $U_{\text{CSWAP}}^{2\otimes2\otimes2}$. VBS$_{i}$ ($i=1,2$) represents variable beam splitter, their rules can be given by Equation  (\ref{eq4}). b) Schematic diagram of the controlled-SWAP gate $U_{\text{CSWAP}}^{2\otimes2\otimes2}$. The left beam displacer (BD) directs horizontally polarized photons to the lower arm and vertically polarized photons to the upper arm, the right beam displacer (BD) recombines photons from two paths with different polarization states. BS$_i$ ($i=1,2,3,4$) denotes the 50:50 beam splitter. P$_\pi$, phase shifter, completes $|H\rangle\rightarrow -|H\rangle$ and $|V\rangle\rightarrow -|V\rangle$ . }
\label{Fig.1}
\end{figure}

It is well known that nonlinear architectures, such as neutral atom, are facing the challenges of inefficiency and impracticality. Linear optical architectures are in general easy to implement with current technology, and the intrinsic probabilistic character can be overcome by exploiting further DOFs or pre-shared multiphoton-entangled states. In the following, we design a simpler scheme for implementing a deterministic three-qubit controlled-SWAP gate, $U_{\text{CSWAP}}^{2\otimes2\otimes2}$. Here the control  and the target qubits are encoded in the polarization and spatial DOFs of a two entangled photon pairs, respectively (see Figure \ref{Fig.1}).

\subsection{Creation of the input state of $U_{\text{CSWAP}}^{2\otimes 2\otimes 2}$}

To quantitatively characterize the gate performance, we first need to generate the input state $|\varphi_{\text{in}}\rangle$ of $U_{\text{CSWAP}}^{2\otimes2\otimes2}$.
Without loss of generality, $|\varphi_{\text{in}}\rangle$ is given by
\begin{eqnarray}                                   \label{eq1}
	\begin{split}
		|\varphi_{\text{in}}\rangle=&(\alpha|0\rangle_c+\beta|1\rangle_c)\otimes(\gamma|0\rangle_{t_1}+\delta|1\rangle_{t_1})\otimes(\mu|0\rangle_{t_2}+\nu|1\rangle_{t_2}),
	\end{split}
\end{eqnarray}
where the complex coefficients $\alpha$, $\beta$, $\gamma$, $\delta$, $\mu$, $\nu$ conform to the normalization condition, i.e., $|\alpha|^2+|\beta|^2=|\gamma|^2+|\delta|^2=|\mu|^2+|\nu|^2=1$. The initial state given by Equation  (\ref{eq1}) can be completed by Figure \ref{Fig.1}(a). Let us action of the scheme, step by step.

In the first step, the entangled photon pair, $\alpha|HH\rangle+\beta|VV\rangle$, occupied spatial modes $b$ and $c$ can be generated by frequency-degenerate type-I spontaneous parametric down-conversion.\textsuperscript{\upcite{type-I}} Here the state of the source is given by
\begin{eqnarray}                                   \label{eq2}
	\begin{split}
		|\varphi_{\text{initial}}\rangle=\alpha \hat{b}_{H}^{\dagger}\hat{c}_{H}^{\dagger}|\text{vac}\rangle+\beta \hat{b}_{V}^{\dagger}\hat{c}_{V}^{\dagger}|\text{vac}\rangle,
	\end{split}
\end{eqnarray}
where $\hat{b}^{\dagger}$ ($\hat{c}^{\dagger}$) denotes the creation operators in spatial mode $b$ ($c$), $H$ ($V$) represents the horizontal (vertical) linear polarization, and $|\text{vac}\rangle$ is the vacuum state.

In the second step, after the spatial modes $b$ and $c$ interferences at two variable beam splitters (VBS$_1$ and VBS$_2$), the state $|\varphi_{\text{initial}}\rangle$ shown in Equation  (\ref{eq2}) becomes
\begin{eqnarray}                                   \label{eq3}
	\begin{split}
		|\varphi_{\text{in}}\rangle=&
		\alpha(\delta\hat{a}_{H}^{\dagger}+\gamma\hat{b}_{H}^{\dagger})(\nu\hat{c}_{H}^{\dagger}+\mu\hat{d}_{H}^{\dagger})|\text{vac}\rangle+
		\beta(\delta\hat{a}_{V}^{\dagger}+\gamma\hat{b}_{V}^{\dagger})(\nu\hat{c}_{V}^{\dagger}+\mu\hat{d}_{V}^{\dagger})|\text{vac}\rangle\\
		=&
		\alpha\delta\nu\hat{a}_{H}^{\dagger}\hat{c}_{H}^{\dagger}|\text{vac}\rangle+
		\alpha\delta\mu\hat{a}_{H}^{\dagger}\hat{d}_{H}^{\dagger}|\text{vac}\rangle+
		\alpha\gamma\nu\hat{b}_{H}^{\dagger}\hat{c}_{H}^{\dagger}|\text{vac}\rangle+
		\alpha\gamma\mu\hat{b}_{H}^{\dagger}\hat{d}_{H}^{\dagger}|\text{vac}\rangle\\
		&+
		\beta\delta\nu\hat{a}_{V}^{\dagger}\hat{c}_{V}^{\dagger}|\text{vac}\rangle+
		\beta\delta\mu\hat{a}_{V}^{\dagger}\hat{d}_{V}^{\dagger}|\text{vac}\rangle+
		\beta\gamma\nu\hat{b}_{V}^{\dagger}\hat{c}_{V}^{\dagger}|\text{vac}\rangle+
		\beta\gamma\mu\hat{b}_{V}^{\dagger}\hat{d}_{V}^{\dagger}|\text{vac}\rangle.
	\end{split}
\end{eqnarray}
Here VBS$_1$ and VBS$_2$ complete the following transformations
\begin{eqnarray}                                   \label{eq4}
	\begin{split}
		&\hat{b}_{H}^{\dagger}\xrightarrow{\text{VBS}_{1}}\delta\hat{a}_{H}^{\dagger}+\gamma\hat{b}_{H}^{\dagger},\quad
		\hat{b}_{V}^{\dagger}\xrightarrow{\text{VBS}_{1}}\delta\hat{a}_{V}^{\dagger}+\gamma\hat{b}_{V}^{\dagger},\\
		&\hat{c}_{H}^{\dagger}\xrightarrow{\text{VBS}_{2}}\nu\hat{c}_{H}^{\dagger}+\mu\hat{d}_{H}^{\dagger},\quad
		\hat{c}_{V}^{\dagger}\xrightarrow{\text{VBS}_{2}}\nu\hat{c}_{V}^{\dagger}+\mu\hat{d}_{V}^{\dagger}.
	\end{split}
\end{eqnarray}
The variable reflectivity ($\delta$ and $\mu$) and transmittance ($\gamma$ and $\nu$) coefficients of the BS have been achieved.\textsuperscript{\cite{VBS}}

Based on Equation  (\ref{eq3}), we can find that if we encode the gate qubits as following
\begin{eqnarray}                                   \label{eq5}
	\begin{split}
		|HH\rangle\equiv|0\rangle_{c},\quad |VV\rangle\equiv|1\rangle_{c},\quad
		|a\rangle\equiv|0\rangle_{t_1},\quad |b\rangle\equiv|1\rangle_{t_1},\quad |c\rangle\equiv|0\rangle_{t_2},\quad |d\rangle\equiv|1\rangle_{t_2}.
	\end{split}
\end{eqnarray}
Equation (\ref{eq1}) is corresponding to
\begin{eqnarray}                                  \label{eq6}
	\begin{split}
		|\varphi_{\text{in}}\rangle=&\alpha\delta\nu|000\rangle_{ct_1t_2}+
		\alpha\delta\mu|001\rangle_{ct_1t_2}+\alpha\gamma\nu|010\rangle_{ct_1t_2}+\alpha\gamma\mu|011\rangle_{ct_1t_2}\\
		&+\beta\delta\nu|100\rangle_{ct_1t_2}+\beta\delta\mu|101\rangle_{ct_1t_2}+\beta\gamma\nu|110\rangle_{ct_1t_2}+\beta\gamma\mu|111\rangle_{ct_1t_2}.
	\end{split}
\end{eqnarray}
Note that the encoding given by Equation  (\ref{eq5}) in our scheme is the opposite of the encoding $|HH\rangle=|0\rangle_1,|VV\rangle=|1\rangle_1,|a\rangle=|1\rangle_2,|b\rangle=|0\rangle_2,|c\rangle=|1\rangle_3,|d\rangle=|0\rangle_3$ in Ref. \cite{Fredkin-Meng}.
And then, the state preparation stage is successfully completed.

\subsection{Scheme for implementing $U_{\text{CSWAP}}^{2\otimes2\otimes2}$}

The schematic diagram for the controlled-SWAP gate $U_{\text{CSWAP}}^{2\otimes2\otimes2}$ is shown Figure \ref{Fig.1}(b), the block (the dashed box) composed of two BSs and one phase shifter with $\theta=\pi$. The two blocks represented in Figure \ref{Fig.1}(b)  perform a spatial-based SWAP operation on the incident photons with the same polarization, i.e.,
\begin{eqnarray}                                  \label{eq7}
	\begin{split}
		\hat{m}_{H}^{\dagger}\hat{n}_{H}^{\dagger}|\text{vac}\rangle\xrightarrow{\text{Blocks}}\hat{n}_{H}^{\dagger}\hat{m}_{H}^{\dagger}|\text{vac}\rangle,\quad
		\hat{m}_{V}^{\dagger}\hat{n}_{V}^{\dagger}|\text{vac}\rangle\xrightarrow{\text{Blocks}}\hat{n}_{V}^{\dagger}\hat{m}_{V}^{\dagger}|\text{vac}\rangle.
	\end{split}
\end{eqnarray}
Here the 50:50 BS  completes the transformations
\begin{eqnarray}                                  \label{eq8}
	\begin{split}
		&\hat{a}_{H}^{\dagger}\xrightarrow{\text{BS}_{1}}\frac{1}{\sqrt{2}}(\hat{a}_{H}^{\dagger}+\hat{c}_{H}^{\dagger}),\quad
		\hat{a}_{V}^{\dagger}\xrightarrow{\text{BS}_{1}}\frac{1}{\sqrt{2}}(\hat{a}_{V}^{\dagger}+\hat{c}_{V}^{\dagger}),\\
		&\hat{c}_{H}^{\dagger}\xrightarrow{\text{BS}_{1}}\frac{1}{\sqrt{2}}(\hat{a}_{H}^{\dagger}-\hat{c}_{H}^{\dagger}),\quad
		\hat{c}_{V}^{\dagger}\xrightarrow{\text{BS}_{1}}\frac{1}{\sqrt{2}}(\hat{a}_{V}^{\dagger}-\hat{c}_{V}^{\dagger}),\\
		&\hat{b}_{H}^{\dagger}\xrightarrow{\text{BS}_{3}}\frac{1}{\sqrt{2}}(\hat{b}_{H}^{\dagger}+\hat{d}_{H}^{\dagger}),\quad
		\hat{b}_{V}^{\dagger}\xrightarrow{\text{BS}_{3}}\frac{1}{\sqrt{2}}(\hat{b}_{V}^{\dagger}+\hat{d}_{V}^{\dagger}),\\
		&\hat{d}_{H}^{\dagger}\xrightarrow{\text{BS}_{3}}\frac{1}{\sqrt{2}}(\hat{b}_{H}^{\dagger}-\hat{d}_{H}^{\dagger}),\quad
		\hat{d}_{V}^{\dagger}\xrightarrow{\text{BS}_{3}}\frac{1}{\sqrt{2}}(\hat{b}_{V}^{\dagger}-\hat{d}_{V}^{\dagger}).
	\end{split}
\end{eqnarray}
The phase shifter orientate at $\theta=\pi$, resulting in
\begin{eqnarray}                                  \label{eq9}
	\begin{split}
		&\hat{a}_{H}^{\dagger}|\text{vac}\rangle\xrightarrow{P_{\pi}}-\hat{a}_{H}^{\dagger}|\text{vac}\rangle,\quad
		\hat{a}_{V}^{\dagger}|\text{vac}\rangle\xrightarrow{P_{\pi}}-\hat{a}_{V}^{\dagger}|\text{vac}\rangle,\\
		&\hat{b}_{H}^{\dagger}|\text{vac}\rangle\xrightarrow{P_{\pi}}-\hat{b}_{H}^{\dagger}|\text{vac}\rangle,\quad
		\hat{b}_{V}^{\dagger}|\text{vac}\rangle\xrightarrow{P_{\pi}}-\hat{b}_{V}^{\dagger}|\text{vac}\rangle.
	\end{split}
\end{eqnarray}

The scheme for controlled-SWAP gate $U_{\text{CSWAP}}^{2\otimes2\otimes2}$ can be expressed by three steps. Firstly, the polarization-dependent calcite beam displacer (BD) orientates
$\{\hat{a}_{V}^{\dagger},\hat{c}_{V}^{\dagger}\}$ to the upper block;
orientates $\{\hat{b}_{V}^{\dagger},\hat{d}_{V}^{\dagger}\}$ to the lower block; orientates $\{\hat{a}_{H}^{\dagger}\hat{c}_{H}^{\dagger}, \hat{a}_{H}^{\dagger}\hat{d}_{H}^{\dagger},
\hat{b}_{H}^{\dagger}\hat{c}_{H}^{\dagger},
\hat{b}_{H}^{\dagger}\hat{d}_{H}^{\dagger}\}$ to the right beam displacer directly.
Subsequently, the two blocks are applied on the incident photons to swap their spatial information.
Lastly, the wave-packet pairs $(\hat{a}_{H}^{\dagger}, \hat{a}_{V}^{\dagger}$),
$(\hat{b}_{H}^{\dagger}, \hat{b}_{V}^{\dagger}$),	$(\hat{c}_{H}^{\dagger}, \hat{c}_{V}^{\dagger}$), and
$(\hat{d}_{H}^{\dagger}, \hat{d}_{V}^{\dagger}$) are rejoined  by beam displacer, respectively. And then, the input state described in Equation  (\ref{eq3}) is transformed into
\begin{eqnarray}                                   \label{eq10}
	\begin{split}
		|\varphi_{\text{out}}\rangle=&
		\alpha\delta\nu\hat{a}_{H}^{\dagger}\hat{c}_{H}^{\dagger}|\text{vac}\rangle+
		\alpha\delta\mu\hat{a}_{H}^{\dagger}\hat{d}_{H}^{\dagger}|\text{vac}\rangle+
		\alpha\gamma\nu\hat{b}_{H}^{\dagger}\hat{c}_{H}^{\dagger}|\text{vac}\rangle+
		\alpha\gamma\mu\hat{b}_{H}^{\dagger}\hat{d}_{H}^{\dagger}|\text{vac}\rangle\\&+
		\beta\delta\nu\hat{a}_{V}^{\dagger}\hat{c}_{V}^{\dagger}|\text{vac}\rangle+
		\beta\delta\mu\hat{b}_{V}^{\dagger}\hat{c}_{V}^{\dagger}|\text{vac}\rangle+
		\beta\gamma\nu\hat{a}_{V}^{\dagger}\hat{d}_{V}^{\dagger}|\text{vac}\rangle+
		\beta\gamma\mu\hat{b}_{V}^{\dagger}\hat{d}_{V}^{\dagger}|\text{vac}\rangle.
	\end{split}
\end{eqnarray}
It corresponds to the output state
\begin{eqnarray}                                   \label{eq11}
	\begin{split}
		|\varphi_{\text{out}}\rangle
		=&\alpha\delta\nu|000\rangle_{ct_1t_2}+\alpha\delta\mu|001\rangle_{ct_1t_2}+\alpha\gamma\nu|010\rangle_{ct_1t_2}+\alpha\gamma\mu|011\rangle_{ct_1t_2}\\
		&+\beta\delta\nu|100\rangle_{ct_1t_2}+\beta\delta\mu|110\rangle_{ct_1t_2}+\beta\gamma\nu|101\rangle_{ct_1t_2}+\beta\gamma\mu|111\rangle_{ct_1t_2}.
	\end{split}
\end{eqnarray}

Therefore, the compact quantum circuit illustrated in Figure \ref{Fig.1}(b) successfully complete an entangled controlled-SWAP gate $U_{\text{CSWAP}}^{2\otimes 2\otimes 2}$ by solely eight linear optics, and the shallow of the circuit is only five.

\section{Implementation of a controlled-SWAP gate $U_{\text{CSWAP}}^{2\otimes 3\otimes 3}$} \label{sec3}

Our approach can be extended to the controlled-SWAP gate on $\mathbb{C}^{2}$$\otimes$$\mathbb{C}^{3}$$\otimes$$\mathbb{C}^{3}$. As the argument made in Section \ref{sec2}, the control and target qubits of such gate $U_{\text{CSWAP}}^{2\otimes 3\otimes 3}$ are encoded as
\begin{eqnarray}                                   \label{eq12}
	\begin{split}
		&|HH\rangle\equiv|0\rangle_{1},\;\; |VV\rangle\equiv|1\rangle_{1},\\
		&|a\rangle\equiv|0\rangle_{2},\;\;\;\quad
		|b\rangle\equiv|1\rangle_{2},\;\;\quad
		|e\rangle\equiv|2\rangle_{2},\\
		&|c\rangle\equiv|0\rangle_{3},\;\;\;\quad
		|d\rangle\equiv|1\rangle_{3},\;\;\quad
		|f\rangle\equiv|2\rangle_{3}.
	\end{split}
\end{eqnarray}

\subsection{Creation of the input state of $U_{\text{CSWAP}}^{2\otimes 3\otimes 3}$}

The state-preparation stage can be completed by using Figure \ref{Fig.2}(a). Firstly, the entangled photon pairs is initially prepared as
\begin{eqnarray}                                   \label{eq13}
	\begin{split}
		|\psi_{\text{initial}}\rangle=\alpha \hat{b}_{H}^{\dagger}\hat{c}_{H}^{\dagger}|\text{vac}\rangle+\beta \hat{b}_{V}^{\dagger}\hat{c}_{V}^{\dagger}|\text{vac}\rangle.
	\end{split}
\end{eqnarray}

Nextly, as shown in Figure \ref{Fig.2}(a), the photons in the state $\hat{b}_{H}^{\dagger}|\text{vac}\rangle$ and $\hat{b}_{V}^{\dagger}|\text{vac}\rangle$ arrives at VBS$_1$ and followed by VBS$_3$, the photons in the state $\hat{c}_{H}^{\dagger}|\text{vac}\rangle$ and $\hat{c}_{V}^{\dagger}|\text{vac}\rangle$ arrives at VBS$_2$ and followed by VBS$_4$. Here $|\psi_{\text{initial}}\rangle$ induced by VBS$_1$ and VBS$_2$ becomes
\begin{eqnarray}                                  \label{eq14}
	\begin{split}
		|\psi_1\rangle= \alpha(\gamma\hat{d}_{H}^{\dagger}+\delta\hat{e}_{H}^{\dagger})(\mu\hat{c}_{H}^{\dagger}+\nu\hat{f}_{H}^{\dagger})|\text{vac}\rangle
		+\beta(\gamma\hat{d}_{V}^{\dagger}+\delta\hat{e}_{V}^{\dagger})(\mu\hat{c}_{V}^{\dagger}+\nu\hat{f}_{V}^{\dagger})|\text{vac}\rangle.
	\end{split}
\end{eqnarray}
${\text{VBS}_{3}}$ and ${\text{VBS}_{4}}$ convert $|\psi_1\rangle$ into
\begin{eqnarray}                                   \label{eq15}
	\begin{split}
		|\psi_{\text{in}}\rangle=&
		\alpha[(\gamma(\varsigma\hat{a}_{H}^{\dagger}+\xi\hat{b}_{H}^{\dagger})+\delta\hat{e}_{H}^{\dagger})(\mu\hat{c}_{H}^{\dagger}+\nu(\tilde{\varsigma}\hat{d}_{H}^{\dagger}+\tilde{\xi}\hat{f}_{H}^{\dagger}))]|\text{vac}\rangle\\
		&+\beta[(\gamma(\varsigma\hat{a}_{V}^{\dagger}+\xi\hat{b}_{V}^{\dagger})+\delta\hat{e}_{V}^{\dagger})(\mu\hat{c}_{V}^{\dagger}+\nu(\tilde{\varsigma}\hat{d}_{V}^{\dagger}+\tilde{\xi}\hat{f}_{V}^{\dagger}))]|\text{vac}\rangle.
	\end{split}
\end{eqnarray}
If we encode the computation qubits as  Equation  (\ref{eq12}),  Equation  (\ref{eq15}) is rewritten as
\begin{eqnarray}                                \label{eq16}
	\begin{split}
		|\psi_{\text{in}}\rangle=&\alpha[\gamma\varsigma(\mu|000\rangle_{123}+\nu\tilde{\varsigma}|001\rangle_{123}+\nu\tilde{\xi}|002\rangle_{123})\\
		&+\gamma\xi(\mu|010\rangle_{123}+\nu\tilde{\varsigma}|011\rangle_{123}+\nu\tilde{\xi}|012\rangle_{123})\\
		&+\delta(\mu|020\rangle_{123}+\nu\tilde{\varsigma}|021\rangle_{123}+\nu\tilde{\xi}|022\rangle_{123})]\\
		&+\beta[\gamma\varsigma(\mu|100\rangle_{123}+\nu\tilde{\varsigma}|101\rangle_{123}+\nu\tilde{\xi}|102\rangle_{123})\\
		&+\gamma\xi(\mu|110\rangle_{123}+\nu\tilde{\varsigma}|111\rangle_{123}+\nu\tilde{\xi}|112\rangle_{123})\\
		&+\delta(\mu|120\rangle_{123}+\nu\tilde{\varsigma}|121\rangle_{123}+\nu\tilde{\xi}|122\rangle_{123})].\\
	\end{split}
\end{eqnarray}
And then, the state-preparation stage is completed.

\subsection{Scheme for implementing $U_{\text{CSWAP}}^{2\otimes3\otimes3}$}

\begin{figure} [htbp]
	\centering
	\includegraphics[width=5.6 cm]{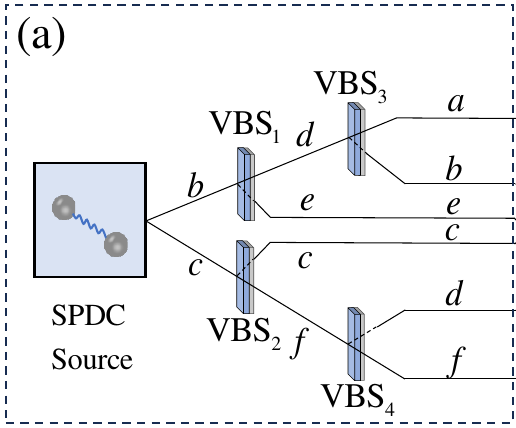}\\	
	\includegraphics[width=12 cm]{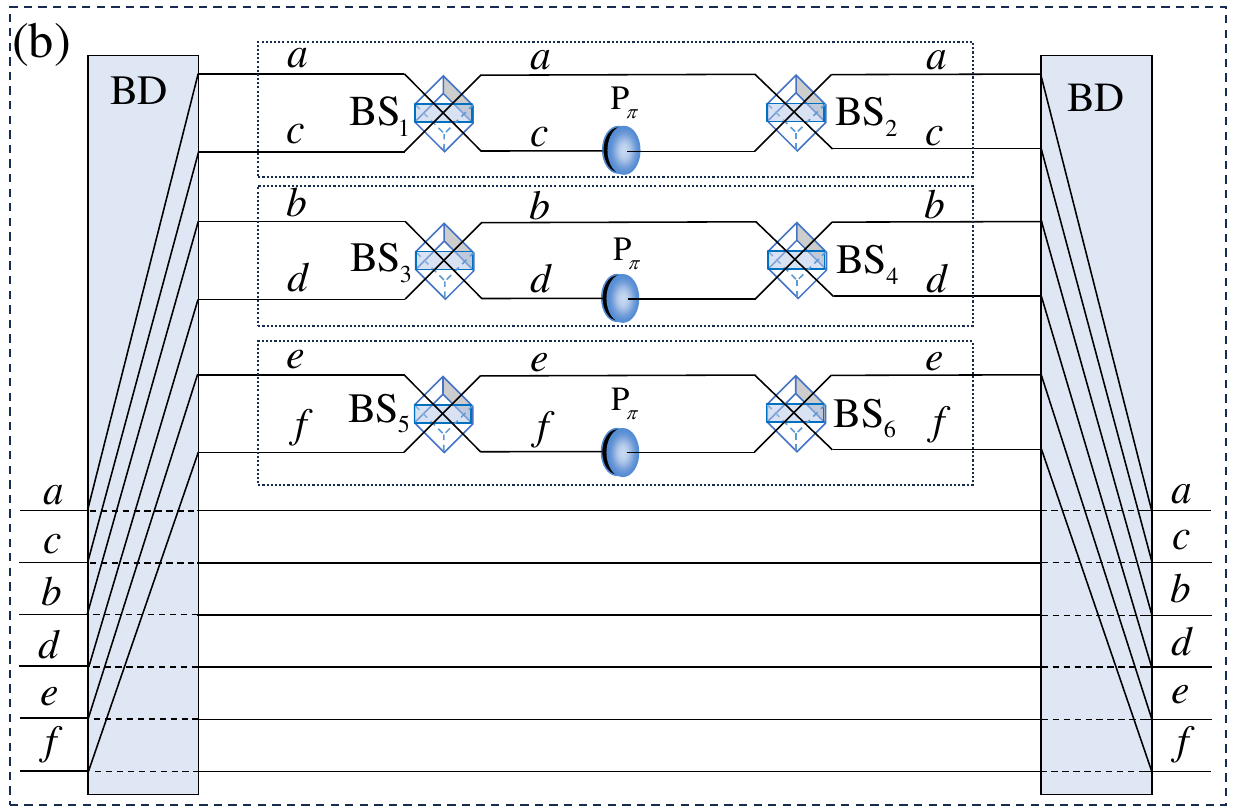}	
	\caption{a) Schematic diagram of the device for state-preparation of controlled-SWAP gate $U_{\text{CSWAP}}^{2\otimes3\otimes3}$. b) Schematic diagram of controlled-SWAP gate $U_{\text{CSWAP}}^{2\otimes3\otimes3}$.}
	\label{Fig.2}
\end{figure}

The linear optical architecture is exhibited in Figure \ref{Fig.2}(b).
Firstly, BD leads $\{\hat{a}_{V}^{\dagger}, \hat{c}_{V}^{\dagger}\}$ to the upper block, leads $\{\hat{b}_{V}^{\dagger}, \hat{d}_{V}^{\dagger}\}$ to the middle block,
and leads $\{\hat{e}_{V}^{\dagger}, \hat{f}_{V}^{\dagger}\}$ to the lower block, and induces $\{\hat{a}_{H}^{\dagger}, \hat{b}_{H}^{\dagger}, \hat{c}_{H}^{\dagger}, \hat{d}_{H}^{\dagger}, \hat{e}_{H}^{\dagger}, \hat{f}_{H}^{\dagger}\}$ bypass the blocks and arrive the right beam displacer directly.

Subsequently, the three blocks acting on the incident photons makes the state given by Equation  (\ref{eq15}) becomes
\begin{eqnarray}                               \label{eq17}
	\begin{split}
		|\psi_{2}\rangle= &\alpha[\gamma\varsigma(\mu\hat{a}_{H}^{\dagger}\hat{c}_{H}^{\dagger}+\nu\tilde{\varsigma}\hat{a}_{H}^{\dagger}\hat{d}_{H}^{\dagger}+\nu\tilde{\xi}\hat{a}_{H}^{\dagger}\hat{f}_{H}^{\dagger})\\
		&+\gamma\xi(\mu\hat{b}_{H}^{\dagger}\hat{c}_{H}^{\dagger}+\nu\tilde{\varsigma}\hat{b}_{H}^{\dagger}\hat{d}_{H}^{\dagger}+\nu\tilde{\xi}\hat{b}_{H}^{\dagger}\hat{f}_{H}^{\dagger})\\
		&+\delta(\mu\hat{e}_{H}^{\dagger}\hat{c}_{H}^{\dagger}+\nu\tilde{\varsigma}\hat{e}_{H}^{\dagger}\hat{d}_{H}^{\dagger}+\nu\tilde{\xi}\hat{e}_{H}^{\dagger}\hat{f}_{H}^{\dagger})]|\text{vac}\rangle\\
		&+\beta[\gamma\varsigma(\mu\hat{c}_{V}^{\dagger}\hat{a}_{V}^{\dagger}+\nu\tilde{\varsigma}\hat{c}_{V}^{\dagger}\hat{b}_{V}^{\dagger}+\nu\tilde{\xi}\hat{c}_{V}^{\dagger}\hat{e}_{V}^{\dagger})\\
		&+\gamma\xi(\mu\hat{d}_{V}^{\dagger}\hat{a}_{V}^{\dagger}+\nu\tilde{\varsigma}\hat{d}_{V}^{\dagger}\hat{b}_{V}^{\dagger}+\nu\tilde{\xi}\hat{d}_{V}^{\dagger}\hat{e}_{V}^{\dagger})\\
		&+\delta(\mu\hat{f}_{V}^{\dagger}\hat{a}_{V}^{\dagger}+\nu\tilde{\varsigma}\hat{f}_{V}^{\dagger}\hat{b}_{V}^{\dagger}+\nu\tilde{\xi}\hat{f}_{V}^{\dagger}\hat{e}_{V}^{\dagger})]|\text{vac}\rangle.\\
	\end{split}
\end{eqnarray}

Lastly,  the components
$\{\hat{a}_{H}^{\dagger}, \hat{a}_{V}^{\dagger}\}$,
$\{\hat{b}_{H}^{\dagger}, \hat{b}_{V}^{\dagger}\}$,
$\{\hat{c}_{H}^{\dagger}, \hat{c}_{V}^{\dagger}\}$,
$\{\hat{d}_{H}^{\dagger}, \hat{d}_{V}^{\dagger}\}$,
$\{\hat{e}_{H}^{\dagger}, \hat{e}_{V}^{\dagger}\}$ and $\{\hat{f}_{H}^{\dagger}, \hat{f}_{V}^{\dagger}\}$
are rejoined by right BD. And then, the state of the composite system becomes
\begin{eqnarray}                           \label{eq18}
	\begin{split}
		|\psi_{\text{out}}\rangle=&\alpha[\gamma\varsigma(\mu\hat{a}_{H}^{\dagger}\hat{c}_{H}^{\dagger}+\nu\tilde{\varsigma}\hat{a}_{H}^{\dagger}\hat{d}_{H}^{\dagger}+\nu\tilde{\xi}\hat{a}_{H}^{\dagger}\hat{f}_{H}^{\dagger})\\
		&+\gamma\xi(\mu\hat{b}_{H}^{\dagger}\hat{c}_{H}^{\dagger}+\nu\tilde{\varsigma}\hat{b}_{H}^{\dagger}\hat{d}_{H}^{\dagger}+\nu\tilde{\xi}\hat{b}_{H}^{\dagger}\hat{f}_{H}^{\dagger})\\
		&+\delta(\mu\hat{e}_{H}^{\dagger}\hat{c}_{H}^{\dagger}+\nu\tilde{\varsigma}\hat{e}_{H}^{\dagger}\hat{d}_{H}^{\dagger}+\nu\tilde{\xi}\hat{e}_{H}^{\dagger}\hat{f}_{H}^{\dagger})]|\text{vac}\rangle\\
		&+\beta[\gamma\varsigma(\mu\hat{a}_{V}^{\dagger}\hat{c}_{V}^{\dagger}+\nu\tilde{\varsigma}\hat{a}_{V}^{\dagger}\hat{d}_{V}^{\dagger}+\nu\tilde{\xi}\hat{a}_{V}^{\dagger}\hat{f}_{V}^{\dagger})\\
		&+\gamma\xi(\mu\hat{b}_{V}^{\dagger}\hat{c}_{V}^{\dagger}+\nu\tilde{\varsigma}\hat{b}_{V}^{\dagger}\hat{d}_{V}^{\dagger}+\nu\tilde{\xi}\hat{b}_{V}^{\dagger}\hat{f}_{V}^{\dagger})\\
		&+\delta(\mu\hat{e}_{V}^{\dagger}\hat{c}_{V}^{\dagger}+\nu\tilde{\varsigma}\hat{e}_{V}^{\dagger}\hat{d}_{V}^{\dagger}+\nu\tilde{\xi}\hat{e}_{V}^{\dagger}\hat{f}_{V}^{\dagger})]|\text{vac}\rangle,
	\end{split}
\end{eqnarray}
which corresponds to the state
\begin{eqnarray}                               \label{eq19}
	\begin{split}
		|\psi_{\text{out}}\rangle=&\alpha[\gamma\varsigma(\mu|000\rangle_{123}+\nu\tilde{\varsigma}|001\rangle_{123}+\nu\tilde{\xi}|002\rangle_{123})\\
		&+\gamma\xi(\mu|010\rangle_{123}+\nu\tilde{\varsigma}|011\rangle_{123}+\nu\tilde{\xi}|012\rangle_{123})\\
		&+\delta(\mu|020\rangle_{123}+\nu\tilde{\varsigma}|021\rangle_{123}+\nu\tilde{\xi}|022\rangle_{123})]\\
		&+\beta[\gamma\varsigma(\mu|100\rangle_{123}+\nu\tilde{\varsigma}|110\rangle_{123}+\nu\tilde{\xi}|120\rangle_{123})\\
		&+\gamma\xi(\mu|101\rangle_{123}+\nu\tilde{\varsigma}|111\rangle_{123}+\nu\tilde{\xi}|121\rangle_{123})\\
		&+\delta(\mu|102\rangle_{123}+\nu\tilde{\varsigma}|112\rangle_{123}+\nu\tilde{\xi}|122\rangle_{123})].\\
	\end{split}
\end{eqnarray}

Based on Equations  (\ref{eq16}-\ref{eq19}), one can see that the entangled-based controlled-SWAP gate $U_{\text{CSWAP}}^{2\otimes 3\otimes 3}$ can be completed by using eleven linear optics, and the shallow of  quantum circuit is still five.

\section{Implementation of a controlled-SWAP gate $U_{\text{CSWAP}}^{2\otimes d\otimes d}$}\label{sec4}

The device in Figure \ref{Fig.2} can be extended to implement the controlled-SWAP gate on $\mathbb{C}^{2}\otimes \mathbb{C}^{d}\otimes \mathbb{C}^{d}$ ($d>3$), see Figure \ref{Fig.3}. Here we notate the general form of the controlled-SWAP gate $U_{\text{CSWAP}}^{2\otimes d\otimes d}$ as
\begin{eqnarray}                               \label{eq20}
	\begin{split}
		U_{\text{CSWAP}}^{2\otimes d\otimes d}=\sum_{i,j=0}^{d-1}\left( |0ij\rangle\langle0ij|+|1ij\rangle\langle1ji|\right),
	\end{split}
\end{eqnarray}
where $i,j=0,1,2,\cdots,d-1$.

\subsection{Creation of the input state of $U_{\text{CSWAP}}^{2\otimes d\otimes d}$}

\begin{figure} [htbp]
	\centering
	\includegraphics[width=7 cm]{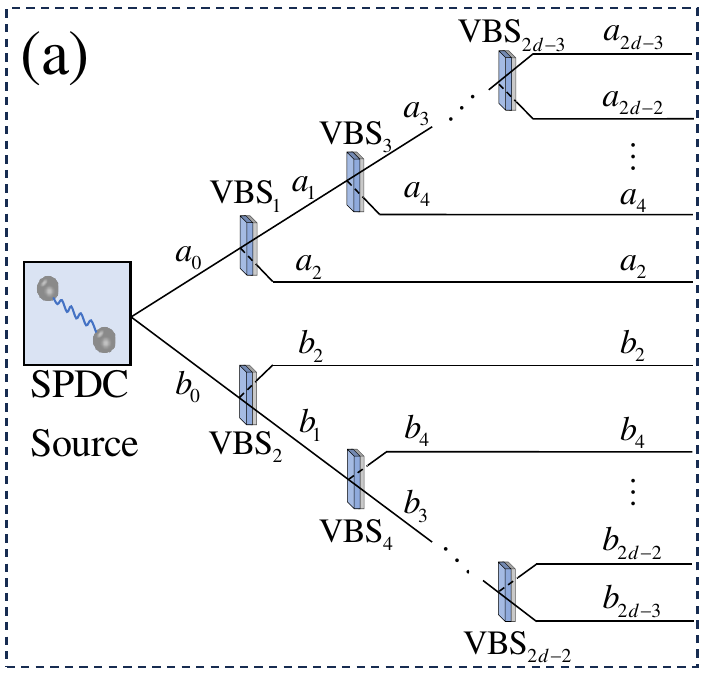}\\
	\includegraphics[width=13 cm]{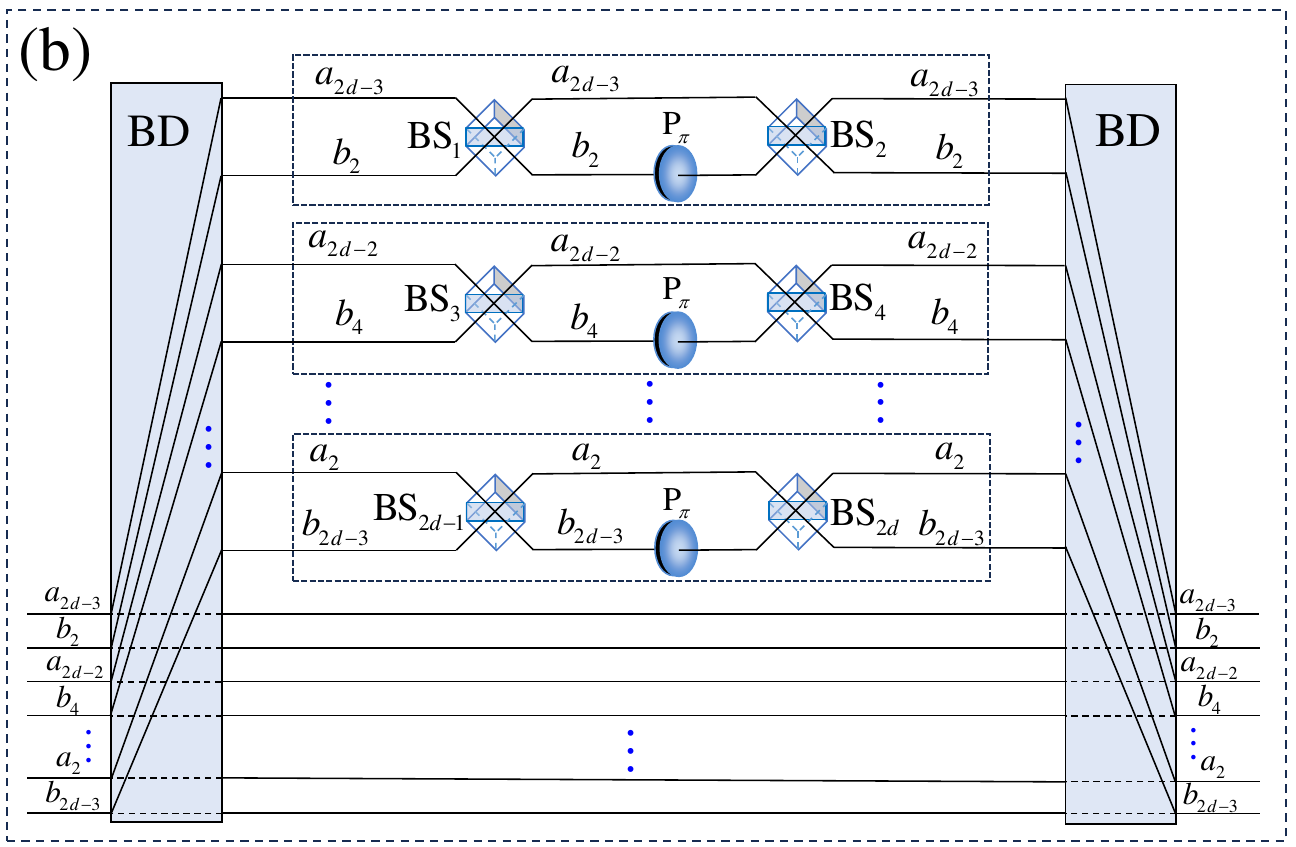}		
	\caption{a) Schematic diagram of the device for state-preparation of controlled-SWAP gate $U_{\text{CSWAP}}^{2\otimes d\otimes d}$. b) Schematic diagram of controlled-SWAP gate $U_{\text{CSWAP}}^{2\otimes d\otimes d}$.}
	\label{Fig.3}
\end{figure}

Similar to the argument made in Section \ref {sec3},
here the Figure \ref{Fig.3}(a) creates the input state of the gate $U_{\text{CSWAP}}^{2\otimes d\otimes d}$. Firstly, the SPDC source produces a polarization-entangled state
\begin{eqnarray}                                   \label{eq21}
	\begin{split}
		|\phi_{\text{initial}}\rangle=\alpha \hat{a}_{0H}^{\dagger}\hat{b}_{0H}^{\dagger}|\text{vac}\rangle+\beta \hat{a}_{0V}^{\dagger}\hat{b}_{0V}^{\dagger}|\text{vac}\rangle,
	\end{split}
\end{eqnarray}
where $\hat{a}_{0}^{\dagger}$ ($\hat{b}_{0}^{\dagger}$) denotes the creation operators in spatial mode ${a}_{0}$ (${b}_{0}$).

Secondly, as shown in Figure \ref{Fig.3}(a), the photons in the state $\hat{a}_{0H}^{\dagger}|\text{vac}\rangle$ and $\hat{a}_{0V}^{\dagger}|\text{vac}\rangle$ arrives at VBS$_1$ and followed by VBS$_3$, the photons in the state $\hat{b}_{0H}^{\dagger}|\text{vac}\rangle$ and $\hat{b}_{0V}^{\dagger}|\text{vac}\rangle$ arrives at VBS$_2$ and followed by VBS$_4$.

Thirdly, growing above steps over and over again, and the triangular array of  $2(d-1)$  VBSs yield the $d-1$ spatial-based basis states $\{ |a_{0}\rangle,|a_{1}\rangle,|a_{2}\rangle,\cdots,|a_{2d-2}\rangle\} $ ($\{ |b_{0}\rangle,|b_{1}\rangle,|b_{2}\rangle,\cdots,|b_{2d-2}\rangle\} $). In other words, after the photon pairs pass through $2(d-1)$ VBSs,	Equation  (\ref{eq21}) becomes
\begin{eqnarray}                                  \label{eq22}
	\begin{split}
		|\phi_{\text{in}}\rangle=\sum_{i,j=0}^{d-1} (\lambda_{0ij}\hat{a}_{iH}^{\dagger}\hat{b}_{jH}^{\dagger}|\text{vac}\rangle+\lambda_{1ij} \hat{a}_{iV}^{\dagger}\hat{b}_{jV}^{\dagger}|\text{vac}\rangle),
	\end{split}
\end{eqnarray}
where the $4(d-1)$ complex parameters generated by $2(d-1)$ VBSs conform to the normalization condition,  i.e., $\sum_{i,j=0}^{d-1}(\left| \lambda_{0ij} \right| ^2+\left|\lambda_{1ij} \right|^2)=1 $.

Similar to the coding method in Section \ref {sec3}, if the control and target qubits of such gate $U_{\text{CSWAP}}^{2\otimes d\otimes d}$ are encoded as
\begin{eqnarray}                                  \label{eq23}
	\begin{split}
		&|HH\rangle\equiv|0\rangle_{1},\;\;\quad |VV\rangle\equiv|1\rangle_{1},\\
		&|a_{2d-3}\rangle\equiv|0\rangle_{2},\;\;\;\; |a_{2d-2}\rangle\equiv|1\rangle_{2},\;\cdots,\; |a_{2}\rangle\equiv|d-1\rangle_{2},\\
		&|b_{2}\rangle\equiv|0\rangle_{3},\;\cdots,\; |b_{2d-3}\rangle\equiv|1\rangle_{3},\;\;\;\qquad |b_{2d-2}\rangle\equiv|d-1\rangle_{3}.	
	\end{split}
\end{eqnarray}
The input state of $U_{\text{CSWAP}}^{2\otimes d\otimes d}$ described in Equation  (\ref{eq22}) corresponds to
\begin{eqnarray}                                   \label{eq24}
	\begin{split}
		|\phi_{\text{in}}\rangle=\sum_{i,j=0}^{d-1} (\lambda_{0ij}|0ij\rangle+\lambda_{1ij}|1ij\rangle).
	\end{split}
\end{eqnarray}
The arbitrary input state of $U_{\text{CSWAP}}^{2\otimes d\otimes d}$ is completed.

\subsection{Scheme for implementing $U_{\text{CSWAP}}^{2\otimes d\otimes d}$}

As shown in Figure \ref{Fig.3}(b), the controlled-SWAP gates $U_{\text{CSWAP}}^{2\otimes d\otimes d}$ can be implemented by our using solely $d$ Blocks which includes  $2d$ BSs and  $d$ phase shifter with $\theta=\pi$.

When the control logic qubit is in the state $|0\rangle_{1}$ corresponding to $|HH\rangle$, BD will orient the photons into the lower arm, the spatial-based target states will not be affected by the blocks (the dashed box), i.e.,
\begin{eqnarray}                                   \label{eq25}
		|0ij\rangle_{123}\xrightarrow{\text{Blocks}}|0ij\rangle_{123},\; (i,j=0,1,\cdots,d-1).
\end{eqnarray}
When the control logic qubit is in the state $|1\rangle_{1}$ corresponding to $|VV\rangle$, BD will orient the photon into the upper arm, the spatial-based target states will  be exchanged by the blocks, i.e.,
\begin{eqnarray}                                   \label{eq26}
	\begin{split}
		|1ij\rangle_{123}\xrightarrow{\text{Blocks}}|1ji\rangle_{123},\; (i,j=0,1,\cdots,d-1).
	\end{split}
\end{eqnarray}
Therefore, the controlled-SWAP gate $U_{\text{CSWAP}}^{2\otimes d\otimes d}$ can be completed by Figure \ref{Fig.3}(b), i.e.,
\begin{eqnarray}                                   \label{eq27}
	\begin{split}
		|\phi_{\text{in}}\rangle\xrightarrow{\text{Blocks}}|\phi_{\text{out}}\rangle=\sum_{i,j=0}^{d-1} (\lambda_{0ij}|0ij\rangle+\lambda_{1ij}|1ji\rangle).
	\end{split}
\end{eqnarray}
Hence the entangled-based controlled-SWAP gate $U_{\text{CSWAP}}^{2\otimes d\otimes d}$ can be completed by using $2+3d$ linear optics. The shallow of the quantum circuit is still 5, and it is polarization independent.

\section{Characterization of the controlled-SWAP gate performance}  \label{sec5}

The imperfect linear optics will reduce the fidelity of the presented controlled-SWAP gates,  as they deviate the real quantum state from the perfect input state. The matrix representations of the imperfect BD and phase shifter in the basis $\{|\hat{\chi}_{H}^{\dagger}\rangle, |\hat{\chi}_{V}^{\dagger}\rangle\}$ are given by
\begin{eqnarray}                            \label{eq28}
	\begin{split}
		U_{\text{beam displacer}}=
		\frac{1}{\sqrt{1+r}}
		\left(
		\begin{array}{cc}
			\cos{\theta}-\sqrt{r}\sin{\theta}	& \sin{\theta}+\sqrt{r}\cos{\theta}  \\
			-\sqrt{r}^{*}\cos{\theta}-\sin{\theta}	& -\sqrt{r}^{*}\sin{\theta}+\cos{\theta} \\
		\end{array}
		\right),
	\end{split}
\end{eqnarray}
\begin{eqnarray}                          \label{eq29}
	\begin{split}
		U_{\text{phase}}=
		\left(
		\begin{array}{cc}
			e^{i(\pi-\Delta\phi)} & 0  \\
			0 & e^{i(\pi-\Delta\phi)} \\
		\end{array}
		\right),
	\end{split}
\end{eqnarray}
where $\theta$ and $r$ are the deviation of mirror mounts and polarization extinction ratios of the beam displacer, respectively.  $\sqrt{r}^{*}$ denotes the
conjugate complex of $\sqrt{r}$. $\phi=\pi-\triangle \phi$ is the real phase shift introduced by the phase shifter. $\chi$ is the spatial mode of the incident photon.

The transformation of the imperfect 50:50 BS acting on the incident photon emitting from two input spatial modes $\chi$ and $\lambda$ is given by
\begin{eqnarray}                                 \label{eq30}
	\begin{split}
		&U_{\text{BS}}
		\left(
		\begin{array}{cc}
			\hat{\chi}_{H}^{\dagger}  \\
			\hat{\lambda}_{H}^{\dagger} \\
		\end{array}
		\right)
		=
		\frac{1}{\sqrt{{\epsilon}^{2}+2\epsilon+2}}
		\left(
		\begin{array}{cc}
			1+\epsilon & 1  \\
			1 & -1-\epsilon \\
		\end{array}
		\right)
		\left(
		\begin{array}{cc}
			\hat{\chi}_{H}^{\dagger}  \\
			\hat{\lambda}_{H}^{\dagger} \\
		\end{array}
		\right),\\
		&U_{\text{BS}}
		\left(
		\begin{array}{cc}
			\hat{\chi}_{V}^{\dagger}  \\
			\hat{\lambda}_{V}^{\dagger} \\
		\end{array}
		\right)
		=
		\frac{1}{\sqrt{{\epsilon}^{2}+2\epsilon+2}}
		\left(
		\begin{array}{cc}
			1+\epsilon & 1  \\
			1 & -1-\epsilon \\
		\end{array}
		\right)
		\left(
		\begin{array}{cc}
			\hat{\chi}_{V}^{\dagger}  \\
			\hat{\lambda}_{V}^{\dagger} \\
		\end{array}
		\right),
	\end{split}
\end{eqnarray}
where $\epsilon$ is the transmission deviation rate of the 50:50 BS, and  two spatial modes $(\chi,\lambda)=(a,c)$ or $(\chi,\lambda)=(b,d)$.

Based on Equation (\ref{eq1}) - Equation (\ref{eq11}), the average fidelity of the presented controlled-SWAP gate  $U_{\text{CSWAP}}^{2\otimes 2\otimes 2}$ can be evaluated as
\begin{eqnarray}                                 \label{eq31}
	\begin{split}
		\bar{F}_{U_\text{CSWAP}^{2\otimes 2\otimes 2}}=
		\frac{1}{(2\pi)^3}\int^{2\pi}_{0}\int^{2\pi}_{0}\int^{2\pi}_{0}|\langle\varphi_{\text{real}}|\varphi_{\text{ideal}}\rangle|^{2}\text{d}x\text{d}y\text{d}z,
	\end{split}
\end{eqnarray}
where the ideal output state $|\varphi_{\text{ideal}}\rangle$ is given by Equation (\ref{eq11}), and $|\varphi_{\text{real}}\rangle$ represents the practical output state with $\alpha=\cos x$, $\beta=\sin x$,  $\gamma=\cos y$, $\delta=\sin y$, $\mu=\cos z$, and $\nu =\sin z$.

\begin{figure}[htbp]
	\centering
	\includegraphics[width=8cm]{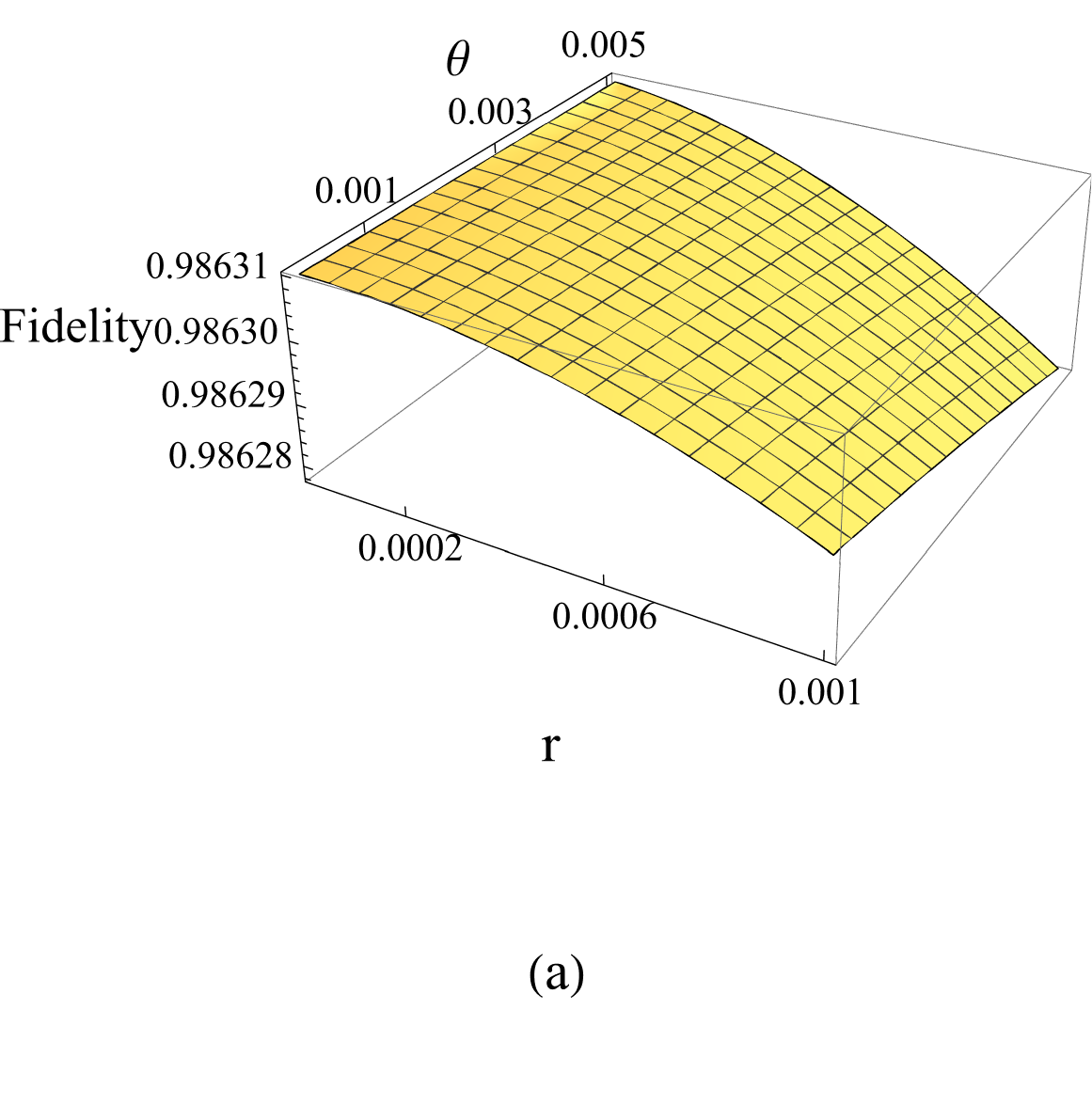}\qquad
	\includegraphics[width=8cm]{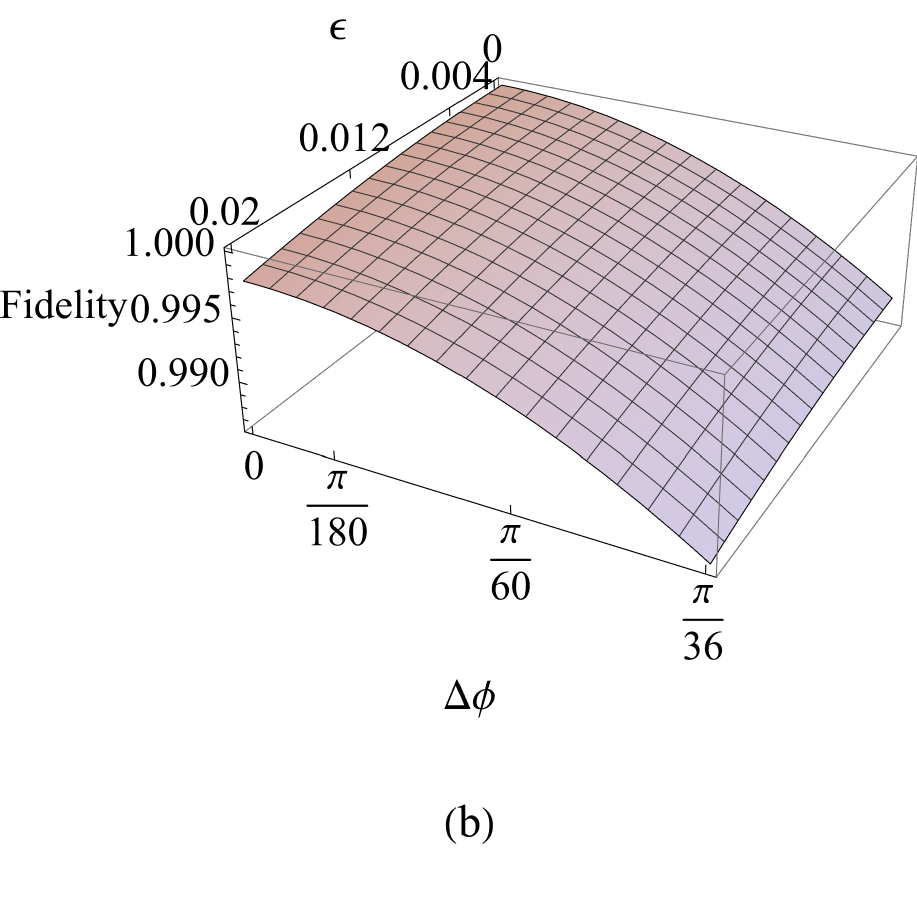}
	\caption{a) The average fidelity $\bar{F}$  of the controlled-SWAP gate $U_{\text{CSWAP}}^{2\otimes 2\otimes 2}$ with $r\in[0,1\times10^{-3}]$ and $\theta\in[0,5\times10^{-3}]$ under realistic conditions $\Delta\phi=\pi/36$ and $\epsilon=0.02$.
b) The average fidelity $\bar{F}$  of the controlled-SWAP gate $U_{\text{CSWAP}}^{2\otimes 2\otimes 2}$ with
		$\epsilon\in[0,0.02]$ and $\Delta\phi\in[0,\pi/36$] under realistic conditions $r=1\times10^{-3}$ and $\theta=5\times10^{-3}$ rad.}
	\label{Fig.4}
\end{figure}

Figure \ref{Fig.4} shows the average fidelity $\bar{F}$ of the controlled-SWAP gate ${U_\text{CSWAP}^{2\otimes 2\otimes 2}}$. Based on Figure \ref{Fig.4}, we one see that when $\Delta\phi=\pi/36$, $\epsilon=0.02$, $r= 1 \times 10^{-3}$ and $\theta= 5 \times 10^{-3}$ rad, the average fidelity approximately $\bar{F}_{U_\text{CSWAP}^{2\otimes 2\otimes 2}} = 0.994$.
When the input state of the controlled-SWAP gate $U_{\text{CSWAP}}^{2\otimes 2\otimes 2}$  are restricted to  $|000\rangle$, $|001\rangle$, $|010\rangle$, and  $|011\rangle$, it is easy to calculate that $\bar{F}_{000} =\bar{F}_{001}=\bar{F}_{010}=\bar{F}_{011}$. Similarly, we can also calculte that
$\bar{F}_{100} =\bar{F}_{111}$ and $\bar{F}_{101}=\bar{F}_{110}$.
The fidelity for input states $|000\rangle$, $|001\rangle$, $|010\rangle$, $|011\rangle$, $|100\rangle$, $|101\rangle$, $|110\rangle$, and $|111\rangle$ is shown in Table \ref{table1}.

\begin{table}[htbp]
	\centering
	\caption{The fidelity of the controlled-SWAP gate $U_{\text{CSWAP}}^{2\otimes 2\otimes 2}$ with $\Delta\phi=\pi/36$, $\epsilon=0.02$, $r= 0.001$ and $\theta= 0.005$ rad. }	\label{table1}
	\begin{tabular}{ccccc}		
		\hline 	
		
		Input state    &  Parameters ($x,y,z$) & Fidelity                   \\			
		\hline	
		$|0ij\rangle,i,j\in\left\lbrace 0,1\right\rbrace $  &  $x=0$, $y,z\in\left\lbrace 0,\pi/2\right\rbrace $ & 0.999 \\	
		$|100\rangle$ &$x=\pi/2,y=z=0$   & 0.986          \\
		$|111\rangle$	 & $x=y=z=\pi/2$ &0.986                \\
		$|101\rangle$  & $x=\pi/2,y=0,z=\pi/2$ & 0.995       \\
		$|110\rangle$	& $x=\pi/2,y=\pi/2,z=0$ &0.995                             \\
		\hline
	\end{tabular}
\end{table}

\begin{figure}[htbp]
\centering
\includegraphics[width=7cm,height=5.5cm]{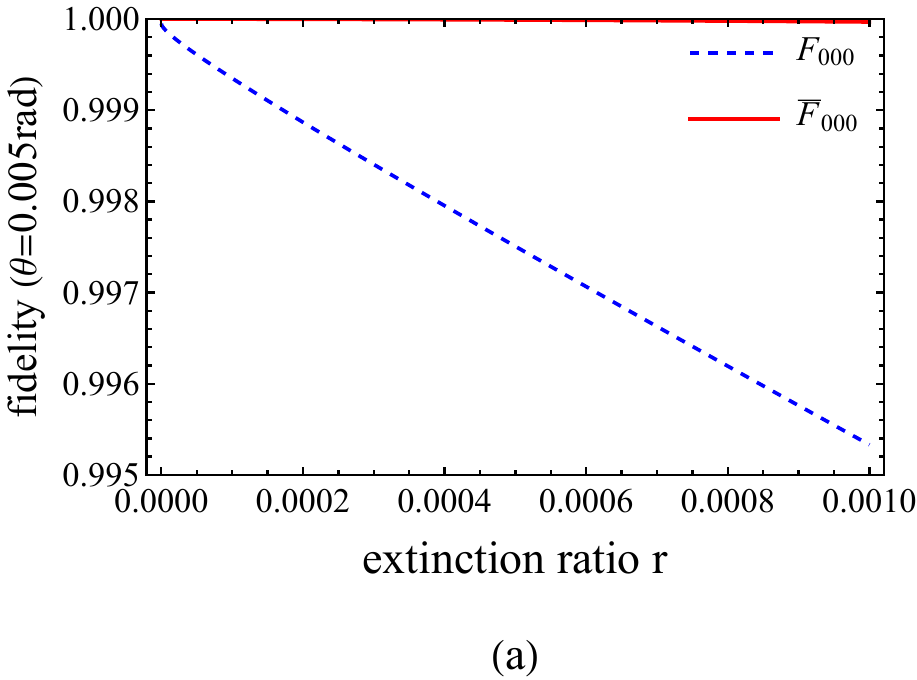}
\qquad\quad
\includegraphics[width=7cm,height=5.5cm]{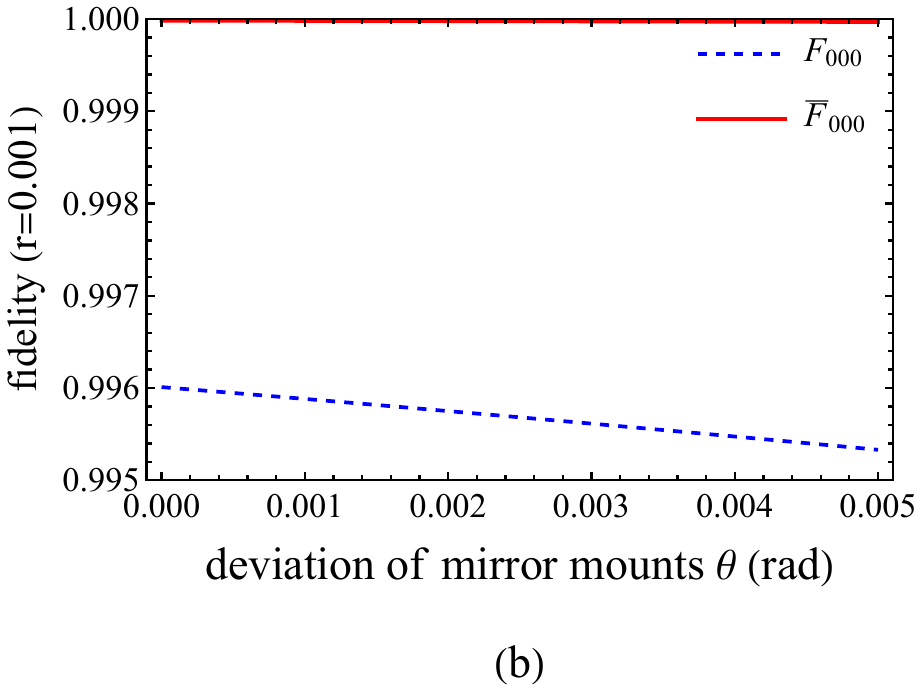}
\caption{a) Fidelities of the controlled-SWAP gate $U_{\text{CSWAP}}^{2\otimes 2\otimes 2}$ for the input state $|000\rangle$ as a function of the extinction ratio $r$ with $\theta=5 \times 10
^{-3}$ rad. b) Fidelities of he controlled-SWAP gate $U_{\text{CSWAP}}^{2\otimes 2\otimes 2}$ for the input state $|000\rangle$ as a function of the deviation of mirror mounts $\theta$ with $r=1\times10
^{-3}$. The solid red curves are for our $\bar{F}_{0ij}$, and the dashed blue curves are for $\mathcal{\bar{F}}_{0ij}$,\upcite{Fredkin-Meng} where $i,j\in\left\lbrace 0,1\right\rbrace $.}
\label{Fig.5}
\end{figure}

\begin{figure}[htbp]
\centering
\includegraphics[width=7cm,height=5.5cm]{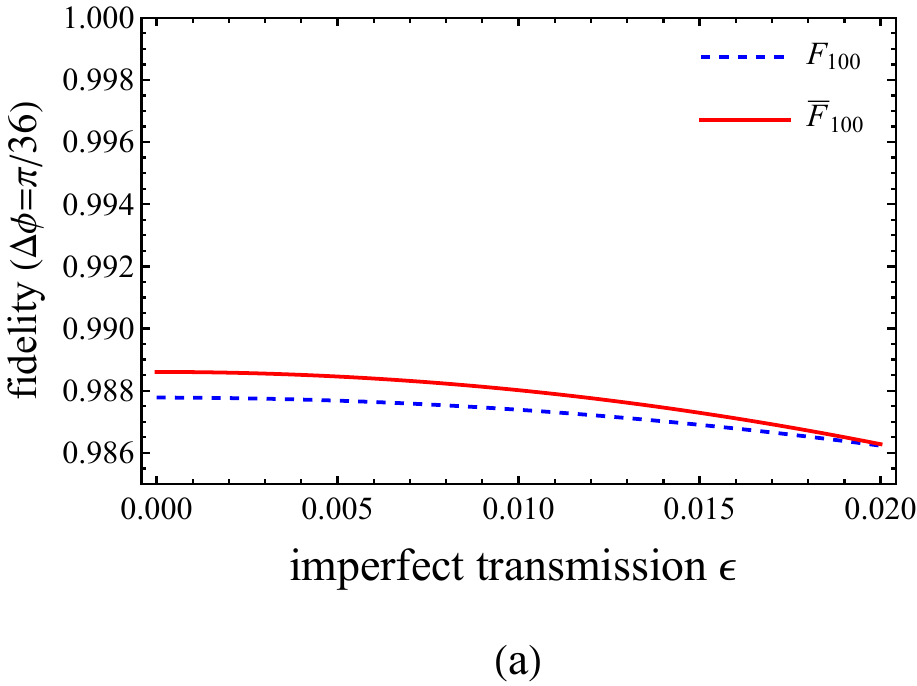}\qquad\quad
\includegraphics[width=7cm,height=5.5cm]{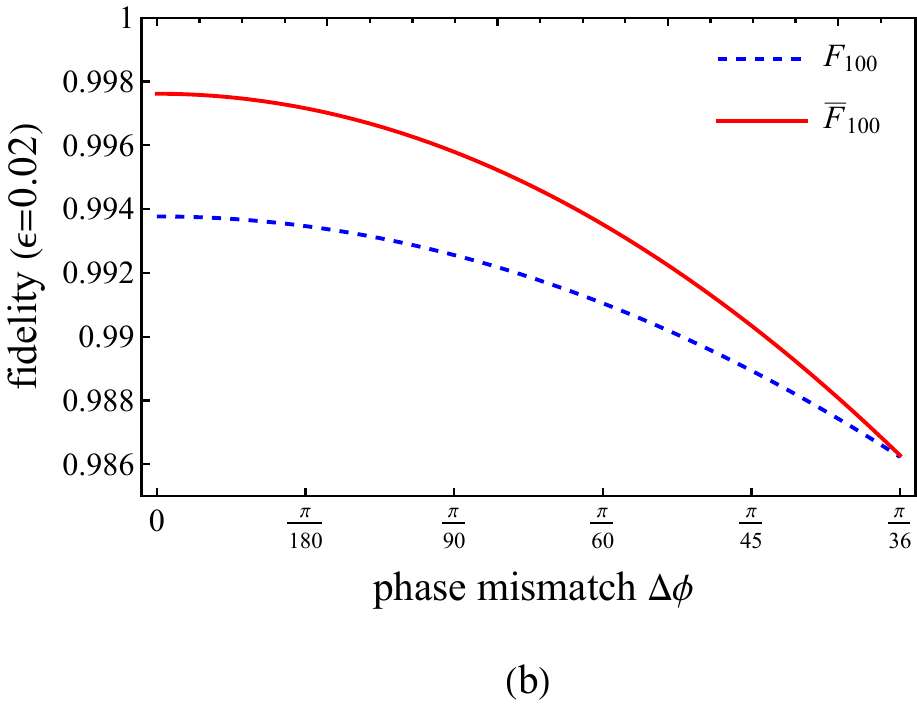}
\caption{a) Fidelities of the controlled-SWAP gate $U_{\text{CSWAP}}^{2\otimes 2\otimes 2}$ for the input state $|100\rangle$ as a function of the imperfect transmission $\epsilon$ with $\Delta\phi=\pi/36$. b) Fidelities of the controlled-SWAP gate $U_{\text{CSWAP}}^{2\otimes 2\otimes 2}$ for the input state $|100\rangle$ as a function of the phase mismatch $\Delta\phi$ with $\epsilon=0.02$. The solid red curves are for our $\bar{F}_{100}$, and the dashed blue curves are for $\mathcal{\bar{F}}_{100}$.\upcite{Fredkin-Meng}}
	\label{Fig.6}
\end{figure}

Figure \ref{Fig.5} shows that $\bar{F}_{0ij}$, $i,j\in\left\lbrace 0,1\right\rbrace $ is much higher than $\mathcal{\bar{F}}_{0ij}$ presented in Ref. \cite{Fredkin-Meng}. For the input state $\bar{F}_{100}$, Figure \ref{Fig.6} shows the fidelity for input state $|100\rangle$ as a function of the imperfect transmission $\epsilon$ and phase mismatch $\Delta\phi$.  Here two imperfect parameters $r= 1\times 10^{-3}$ and $\theta= 5 \times 10^{-3}$ rad are taken.

\begin{table}[htbp]
	\centering
	\caption{Comparison of controlled-SWAP gates with the gates reported in Ref. \cite{Fredkin-Meng}.}	\label{table2}
	\begin{tabular}{ccccc}		
		\hline 	
		&    &  Number of Linear   &      \\
		&   Study     &   optic elements   &    Depth  \\
		
		\hline
		
		$U_{\text{CSWAP}}^{2\otimes2\otimes2}$ & Ref. \cite{Fredkin-Meng}&  14 & 11 \\
		&This work & 8 & 5 	\\		
		
		$U_{\text{CSWAP}}^{2\otimes3\otimes3}$& Ref. \cite{Fredkin-Meng} &  none & none\\
		&This work & 11 & 5&  \\
		
		$U_{\text{CSWAP}}^{2\otimes d\otimes d}$ & Ref. \cite{Fredkin-Meng} &  none & none \\
		&This work &$2+3d$&5\\
		\hline 		
	\end{tabular}
\end{table}

\section{Conclusion}\label{sec6}

Quantum gates are essential for quantum computation, and play an important role in some quantum communication tasks. Hence, designing quantum gates as simple as possible, as shallow as possible, and using as few qubits as possible is crucial in the filed of quantum information processing. High-dimension quantum gate is fascinating quantum resources for its outstanding merits, and it benefit a wide range of application.

In this paper, we present a family of controlled-SWAP gates $U_{\text{CSWAP}}^{2\otimes d\otimes d}$ by using linear optics, where $d\geq 2$. The controlled qubits and the target qubits (qudit) occupy the polarization and spatial DOF of an entangled photon pair. As shown in Table \ref{table2}, in contrast to the scheme for implementing $U_{\text{CSWAP}}^{2\otimes 2\otimes 3}$ presented in Ref. \cite{Fredkin-Meng},  we reduced the number of linear optics from 14 to 8, and compress the circuit depth from 11 to 5. Additionally, we extended the approach to $\mathbb{C}^{2}\otimes \mathbb{C}^{d}\otimes \mathbb{C}^{d}$, and  $U_{\text{CSWAP}}^{2\otimes d\otimes d}$ is constructed by using  2 BDs, $2d$ 50:50 BSs, $d$ phase shifters. Note that the depth of the such circuit is 5, and it is polarization independent.

The evaluation of the gate performance indicates that the presented simplified schemes can be achieved with high fidelity. Moreover, entanglement photon pair makes the gate suitable for deterministic quantum computation in linear optic architecture. The entangled photon pair has been successful achieved in experiment, and we successful extend the target sate to arbitrary high-dimension. Hence, our economical schemes may be potential for high-dimension quantum processing.

\medskip
\section*{Acknowledgment} 
This work is supported by the National Natural Science Foundation of China under Grant No. 62371038 and the Tian-jin Natural Science Foundation under Grant No. 23JCQNJC00560.

\medskip
\section*{Conflict of Interest}
The authors declare no conflict of interest.

\medskip
\section*{Data Availability Statement}
The data that support the findings of this study are available from the corresponding author upon reasonable request.

\end{document}